\documentclass[aps,prc,twocolumn,times,graphicx,tighten,showpacs]{revtex4}
\usepackage[dvips]{graphicx}
\usepackage[dvips]{color}

\newcommand{\bea}{\begin{eqnarray}}
\newcommand{\eea}{\end{eqnarray}}
\newcommand{\be}{\begin{equation}}
\newcommand{\ee}{\end{equation}}
\newcommand{\np}{{\bf p}}
\newcommand{\hp}{\widehat{\bf p}}

\newcommand{\nh}{{\bf h}}

\newcommand{\nq}{{\bf q}}

\begin{document}

\title{
Angular distribution in two-particle emission
induced by neutrinos and electrons
}

\author{
I. Ruiz Simo$^a$,
C. Albertus$^a$,
J.E. Amaro$^a$,
M.B. Barbaro$^b$,
J.A. Caballero$^c$,
T.W. Donnelly$^d$,
}

\affiliation{$^a$Departamento de F\'{\i}sica At\'omica, Molecular y Nuclear,
and Instituto de F\'{\i}sica Te\'orica y Computacional Carlos I,
Universidad de Granada, Granada 18071, Spain}

\affiliation{$^b$Dipartimento di Fisica, Universit\`a di Torino and
  INFN, Sezione di Torino, Via P. Giuria 1, 10125 Torino, Italy}

\affiliation{$^c$Departamento de F\'{\i}sica At\'omica, Molecular y Nuclear,
Universidad de Sevilla, Apdo.1065, 41080 Sevilla, Spain}

\affiliation{$^d$Center for Theoretical Physics, Laboratory for Nuclear
  Science and Department of Physics, Massachusetts Institute of Technology,
  Cambridge, MA 02139, USA}

\date{\today}


\begin{abstract}
The angular distribution of the phase space arising in two-particle
emission reactions induced by electrons and neutrinos is computed in
the laboratory (Lab) system by boosting the isotropic distribution in
the center of mass (CM) system used in Monte Carlo generators. The Lab
distribution has a singularity for some angular values, coming from
the Jacobian of the angular transformation between CM and Lab
systems. We recover the formula we obtained in a previous calculation
for the Lab angular distribution. This is in accordance with the Monte
Carlo method used to generate two-particle events for neutrino
scattering~\cite{Sob12}. Inversely, by performing the transformation
to the CM system, it can be shown that the phase-space function, which
is proportional to the two particle-two hole (2p-2h) hadronic tensor
for a constant current operator, can be computed analytically in the
frozen nucleon 
 approximation,
if Pauli blocking is absent. The results in the CM frame confirm our
previous work done using an alternative approach in the Lab frame. The
possibilities of using this method to compute the hadronic tensor by a
boost to the CM system are analyzed.

\end{abstract}


\maketitle

\section{Introduction}

Multinucleon emission by electroweak probes is of much interest
nowadays~\cite{Gal11,For12,Mor12,Alv14}. Evidence of its presence in
the quasielastic (QE) peak region has been emphasized in the analysis
of recent neutrino and antineutrino scattering
experiments~\cite{Agu10,Agu13,Fio13,Abe13}.  This has been confirmed
by theoretical calculations~\cite{Mar09,Nie11,Ama11,Lal12}, including
in the dynamics various nuclear effects such as meson-exchange
currents (MEC) with and without $\Delta$-isobar excitations,
final-state interactions (FSI), short-range correlations (SRC), the
random-phase approximation (RPA), effective interactions, {\it etc.}
These ingredients lead to discrepancies between the theoretical
predictions, and these need to be clarified in order to reduce the
systematic uncertainties in neutrino data
analyses~\cite{Gra13,Mar13b,Mar14,Ama12}.

The implementation of two-nucleon ejection in Monte Carlo (MC)
neutrino event generators requires an algorithm to generate events of
two-nucleon final states from given values of momentum and energy
transfer. The standard way to proceed, followed in
\cite{Sob12,GENIE,Katori13},
is to select two nucleons from the Fermi sea, invoke energy-momentum
conservation and compute the four-momentum of the final two-nucleon
state (selecting two nucleon momenta in the final state). In the CM
frame one assumes that 
the two final nucleons move back-to-back with the same
given energy and opposite momentum. The emission angles are chosen
assuming an isotropic distribution in the CM. Once the final momenta
are given, a boost is performed to the Lab system to obtain the
momenta of the two ejected nucleons in this frame; these are then
further propagated in the MC cascade model.

We have recently studied the angular distribution in the Lab frame
corresponding to two-particle (2p) emission in the frozen nucleon
approximation~\cite{Rui14}, where the two nucleons are initially at
rest. This distribution appears in the phase-space integration of
the inclusive hadronic tensor in the 2p-2h channel. We found that
the angular distribution has singularities coming from the Jacobian
obtained by integration of the Dirac delta function of energy
conservation, where a denominator appears that can be zero for some
angles. This behaviour is due to the fact that for a fixed pair of
hole momenta $\nh_1,~\nh_2$, and for given momentum transfer, $q$,
and emission angle $\theta'_1$ of the first particle, there are two
solutions for the momentum of the ejected nucleon $p'_1$ that are
compatible with energy conservation. For a given value of the energy
transfer $\omega$, these two solutions collapse into only one for
the maximum allowed emission angle. For this angle there is a
minimum in the 2p-2h excitation energy, $E_{ex}$, as a function of
$p'_1$, and therefore the derivative that appears in the denominator
of the Jacobian is zero: $dE_{ex}/dp'_1=0$.

In \cite{Rui14} we showed that the divergence of the angular distribution
in the Lab system is of the type $\int_0^1 f(x) dx/\sqrt{x}$. Hence
it is integrable around zero, and we gave an analytic formula for
the integral around the divergence. The interest of the detailed study
of the angular integral was to reduce the CPU time in the calculation
of the hadronic tensor for inclusive neutrino scattering. Here a 7D
integral appears that has to be computed in a reasonable time in
order to use it to predict flux integrated neutrino cross sections,
where one additional integration is needed.

In this paper we show that the isotropic angular distribution in the
CM frame, as the one used in Monte Carlo generators~\cite{NuWro},
corresponds exactly to the angular distribution obtained by us in the
Lab system after integration of the Dirac delta function of
energy. Although this correspondence seems to be evident, in practice
it is not so obvious because in Monte Carlo generators no integration
of a delta function of energy is explicitly performed, or at least no
Jacobian is present in the algorithm to select the emission
angle~\cite{Sob12}.  That means that the phase-space angular
distribution in the Monte Carlo codes is known except for a
normalization factor. Besides it was not evident earlier 
why the divergence in
the angular distribution appears in the Lab system from a constant
distribution in the CM and how it can be handled by the Monte Carlo
procedure.

Furthermore, we also show that upon performing the phase-space integral
in the CM system one finds that the result is analytic if there is no Pauli
blocking, and we give a simple formula for it in the frozen nucleon
approximation. This integration method in the CM frame
provides an alternative way to compute the hadronic tensor
in neutrino and electron scattering.

The interest of the present study is directly linked to the
reliability of the frozen nucleon approximation to get sensible
results for intermediate to high momentum and energy transfers. This
was already applied to a preliminary evaluation of the hadronic tensor
in the case of the seagull current.  Moreover the frozen nucleon
approximation is the leading term if the current is expanded in powers
of (h1,h2) around (0,0).  An integral over the emission angle remains
to be performed. Under the assumption that the dependence of the
elementary hadronic tensor on the emission angle is soft, one could
factorize it out of the integral, evaluating it for some average
angle, say $(\theta_{Max}+\theta_{Min})/2$, times the phase-space
integral.  In fact, the strong dependence of the electroweak matrix
elements comes from the $(q,\omega)$ dependence of the electroweak
form factor and not from the angular dependence for fixed
$(q,\omega)$. The validity of these assumptions will be verified in a
coming paper where the angular dependence of the elementary hadronic
tensor will be studied.

In Section II we present a detailed study of the general formalism
with explicit evaluation of the phase space and discussions on how to
perform explicitly the boost between the two reference frames, Lab and
CM.  We introduce all of the variables required to analyze the 2p-2h
problem and make contact with the frozen nucleon approximation where
the calculations can be done in a straightforward way. Importantly, we
show that these ideas can be incorporated into fully relativistic
2p-2h analyses of neutrino reactions. In Section III we summarize our
basic findings and point out the main issues to be considered in
future work, {\it i.e.,} in any approach that attempts to take into
account two-nucleon ejection effects in lepton scattering reactions.

\section{Formalism}

\subsection{Lab frame}

The starting point is the 2p-2h hadronic tensor
for neutrino and electron scattering in the Lab system,
given in the Fermi gas by
\begin{eqnarray}
W^{\mu\nu}_{2p-2h}
&=&
\frac{V}{(2\pi)^9}\int
d^3p'_1
d^3h_1
d^3h_2
\frac{m_N^4}{E_1E_2E'_1E'_2}
\nonumber \\
&&
r^{\mu\nu}(\np'_1,\np'_2,\nh_1,\nh_2)
\delta(E'_1+E'_2-E_1-E_2-\omega)
\nonumber\\
&&
\Theta(p'_1,p'_2,h_1,h_2) \,,
\label{hadronic2}
\end{eqnarray}
where $Q^\mu=(\omega,\nq)$ is the four momentum transfer,
$m_N$ is the nucleon mass, and $V$ is the volume of the system.
The four-momenta of the final particles and holes are
$P'_i=(E'_i,\np'_i)$, and $H_i=(E_i,\nh_i)$, respectively.
Momentum conservation implies  $\bf p'_2= h_1+h_2+q-p'_1$.
The initial Fermi gas ground state and Pauli blocking imply that
$h_i<k_F$, and  $p'_i>k_F$. These conditions
are included in the $\Theta$ function,
defined as the product of step functions
\begin{eqnarray}
\Theta(p'_1,p'_2,h_1,h_2)
&=&
\theta(p'_2-k_F)
\theta(p'_1-k_F)
\nonumber\\
&\times&
\theta(k_F-h_1)
\theta(k_F-h_2) \,.
\end{eqnarray}
The function  $r^{\mu\nu}(\np'_1,\np'_2,\nh_1,\nh_2)$ is the
hadronic tensor for the elementary transition of a nucleon pair
with the given initial and final momenta, 
summed over spin and isospin~\cite{Rui14}.

We choose the ${\bf q}$ direction to be along the $z$-axis. Then the above
integral is reduced to 7 dimensions. First there is a global rotational
symmetry over one of the azimuthal angles. We choose $\phi'_1=0$ and
multiply by a factor $2\pi$. Furthermore, the energy delta function
enables an analytic integration over $p'_1$.
This 7D integral has to be performed numerically~\cite{DePace03,Ama10a}.
Under some approximations~\cite{Donnelly:1978xa,Van80,Alb84,Gil97}
the number of dimensions can be further reduced,
but this cannot be done in the fully relativistic calculation.

In a previous paper~\cite{Rui14} we compared different methods to
evaluate the above integral numerically. In particular we studied the
special case of the phase-space function $F(q,\omega)$, obtained by
using a constant elementary tensor $r^{\mu\nu}=1$ (independent of the
kinematics), defined, except for a factor $V/(2\pi)^9$, as
\begin{eqnarray}
F(q,\omega)
&\equiv&
\int
d^3p'_1
d^3h_1
d^3h_2
\frac{m_N^4}{E_1E_2E'_1E'_2}
\nonumber \\
&&
\delta(E'_1+E'_2-E_1-E_2-\omega)
\Theta(p'_1,p'_2,h_1,h_2)
\nonumber\\
\label{phase}
\end{eqnarray}
with $\bf p'_2= h_1+h_2+q-p'_1$.

For fixed hole momenta, the energy of the two final particles is
\begin{equation}
E'=E'_1+E'_2=\sqrt{p'_1{}^2+m_N^2}+\sqrt{(\np'-\np'_1)^2+m_N^2} \,,
\end{equation}
where
\begin{equation}
\np'=\nh_1+\nh_2+\nq
\end{equation}
is the final momentum of the pair.
For fixed emission angle $\theta'_1$, we integrate over $p'_1$
changing to the variable $E'$.
By differentiation we arrive at the following Jacobian
[note that the Jacobian of \cite{Lal12} agrees
with Eq. (\ref{jacobiano2})]
\begin{equation}
\left|\frac{dp'_1}{dE'}\right|
=\left| \frac{p'_1}{E'_1}-\frac{\np'_2\cdot\hp'_1}{E'_2} \right|^{-1}
\label{jacobiano2}
\end{equation}
with $\hp'_1\equiv \np'_1/p'_1$.
Now integration of the Dirac delta function of energy gives $E'=E_1+E_2+\omega$
and the phase-space function becomes
\begin{eqnarray}
F(q,\omega)
&=&
2\pi
\int
d^3h_1
d^3h_2
d\theta'_1\sin\theta'_1
\frac{m_N^4}{E_1E_2}
\label{integral7Drel}\\
&\times &
\sum_{\alpha=\pm}
\left.
\frac{p'_1{}^2}
{\left| \frac{p'_1}{E'_1}-\frac{\np'_2\cdot\hp'_1}{E'_2} \right|}
\frac{ \Theta(p'_1,p'_2,h_1,h_2) }{ E'_1E'_2 }
\right|_{p'_1= p'_1{}^{(\alpha)}} \,,
\nonumber
\end{eqnarray}
where the sum inside the integral runs over the two solutions
$p'_1{}^{(\pm)}$ of the energy conservation equation which is
quadratic in $p'_1$. The explicit expressions of
the two solutions are given in~\cite{Rui14}.

In this paper we are interested in the angular dependence of the integrand.
We define the angular distribution function
for fixed values of $(q,\omega,\nh_1, \nh_2)$ as
\begin{eqnarray}
\Phi(\theta'_1)
&=&
 \sin\theta'_1 \int p'_1{}^2 dp'_1 \delta(E_1+E_2+\omega-E'_1-E'_2)
\nonumber\\
&&\times
\Theta(p'_1,p'_2,h_1,h_2)
 \frac{m_N^4}{E_1E_2E'_1E'_2}
\nonumber\\
&=&
\sum_{\alpha=\pm}
\left.
\frac{  m_N^4\sin\theta'_1 p'_1{}^2 \Theta(p'_1,p'_2,h_1,h_2)  }
{  E_1E_2E'_1E'_2
   \left| \frac{p'_1}{E'_1}-\frac{\np'_2\cdot\hp'_1}{E'_2} \right|}
\right|_{p'_1= p'_1{}^{(\alpha)}}
\nonumber\\
&\equiv& \Phi_+(\theta'_1) + \Phi_-(\theta'_1) \, ,
\label{angular}
\end{eqnarray}
where $\Phi_{\pm}(\theta'_1)$ correspond to the two terms of the sum.
Once more $\np'_2=\nh_1+\nh_2+\nq-\np'_1$.
The function $\Phi(\theta'_1)$ thus measures the distribution of final
nucleons as a function of the angle $\theta'_1$. Note that this
function is computed analytically in the Lab system, given as a sum
over the two solutions of the energy conservation condition. Thus there are
really two distributions corresponding to the two possible energies of
final particles for a given emission angle. The angular distribution
is referred to the first particle. The second one is
determined by energy-momentum conservation.

In \cite{Rui14} it was shown that the angular distribution 
in Eq.~(\ref{angular}) has divergences for some angles where the
denominator coming from the Jacobian is zero. Examples were given in
the frozen nucleon approximation. It was also shown that the
divergence is integrable, and an analytic formula was given for the
integral over $\theta'_1$ around the divergence. The integral in the
remaining intervals was performed numerically.

\subsection{Boost from the CM frame}

In Monte Carlo event generators the angular distribution is
obtained from an isotropic distribution in the CM frame, and then transformed
back to the Lab system. Here we show that our distribution is
recovered except for a normalization constant that we determine.

First we fix the kinematics of $(q,\omega,\nh_1, \nh_2)$. To simplify
our formalism, we consider the particular case of the frozen nucleon
approximation, {\it i.e.,} $h_1=h_2=0$. The general case can be done
similarly. The frozen nucleon approximation has the advantage that the total
final momentum is equal to $\np'= \nq$ and hence the CM frame moves
in the z-direction (``upwards''). Therefore, the $x,y$ components are
invariant under the boost from the CM to the Lab frames. In
\cite{Rui14} it was shown that the frozen nucleon approximation gives an
accurate representation of the total phase-space function, so one
expects the angular distribution in the frozen nucleon approximation
to be representative of the general case.

Doubly-primed variables refer to the CM system.
The total final momentum is
\begin{equation}
\np''= \np_1''+\np_2''=0 \, ,
\end{equation}
and the total final
energy $E''$ is determined by invariance of the squared four momentum
\begin{equation}
E''= \sqrt{E'^2-{p'}^2} \, ,
\end{equation}
where $(E',\np')=(2m_N+\omega,\nq)$
are the final energy and momentum in the Lab frame.

In the CM frame the two final nucleons are assumed
 to go back-to-back with the same momentum
and with the same energy
\begin{equation}
E''_1=E''_2= \frac{E''}{2} = \frac12 \sqrt{E'^2-{p'}^2} \, .
\end{equation}
The condition $E''_1>m_N$ restricts the allowed $(\omega,q)$ region
where the two-nucleon emission is possible.

Let $\theta''_1$ be the emission angle corresponding to the first
particle. To obtain the nucleon momentum in the Lab system we perform
a boost of the four vector $(P''_1)^\mu= (E''_1,\np''_1)$ back to the
Lab frame, that is moving downward the $z$-axis with dimensionless velocity $v$,
where this is the velocity of the CM system with respect to the Lab system, 
given by
\begin{equation}
v= \frac{p'}{E'} \, .
\end{equation}
The boost transformation of the $(0,z)$ four-vector components
is given by a $2\times 2$ Lorentz matrix equation
\begin{equation}
\left(
\begin{array}{c} E'_1 \\ p'_{1z} \end{array}
\right)
=
\gamma \left(
\begin{array}{cc} 1 & v \\ v & 1 \end{array}
\right)
\left(
\begin{array}{c} E''_1 \\ p''_{1z} \end{array}
\right) \, ,
\end{equation}
where $\gamma\equiv 1/\sqrt{1-v^2}$. From here we get
\begin{eqnarray}
E'_1 &=& \gamma(E''_1+v p''_1\cos\theta''_1) \\
p'_{1}\cos\theta'_1 &=& \gamma(v E''_1+p''_1\cos\theta''_1) \, .
\end{eqnarray}
Therefore the momentum and angle in the Lab system are
\begin{eqnarray}
p'_1 &=& \sqrt{\gamma^2(E''_1+v p''_1\cos\theta''_1)^2-m_N^2}
\label{plab}\\
\cos\theta'_1 &=&
\frac{\gamma(v E''_1+p''_1\cos\theta''_1)}
{\sqrt{\gamma^2(E''_1+v p''_1\cos\theta''_1)^2-m_N^2}} \, .
\label{coslab}
\end{eqnarray}

\begin{figure}
\includegraphics[width=8cm, bb=130 240 430 780]{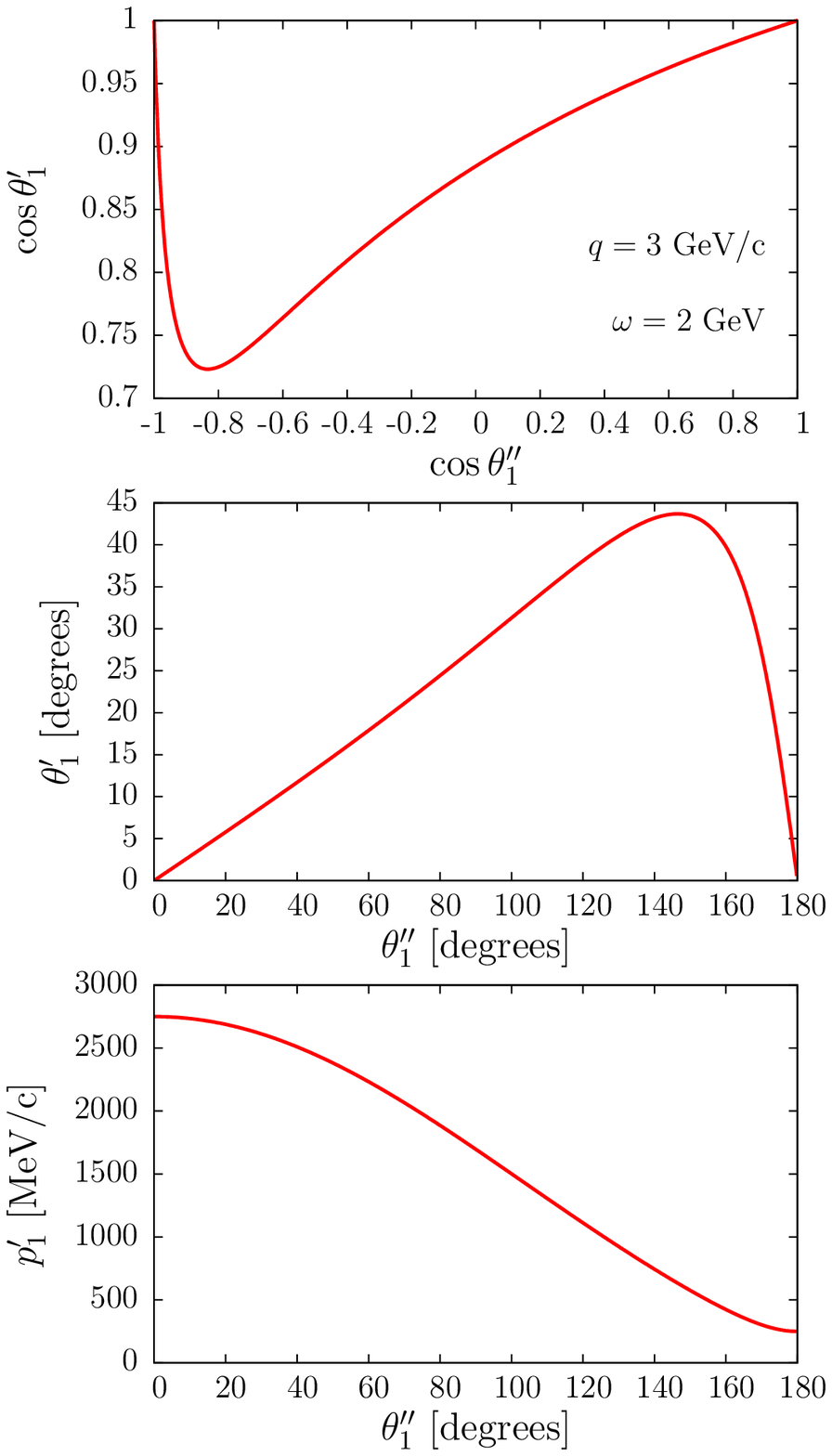}\caption{
\label{fig1}
(Color online) Lab magnitudes as a function of CM magnitudes.  The
momentum and energy transfer are $q=3$ GeV/c, and $\omega=2$ GeV. Top
panel: $\cos\theta'_1$ versus $\cos\theta''_1$. Middle panel:
$\theta'_1$ versus $\theta''_1$. Bottom panel: $p'_1$ versus
$\theta''_1$.  }
\end{figure}

In Fig.~\ref{fig1} we show the Lab emission angle as a function of the
CM angle for momentum and energy transfers: $q=3$ GeV/c and $\omega=2$ GeV. We
choose in this case a high value of the momentum transfer to avoid
effects linked to Pauli blocking. The $\omega$ value is close to the QE
peak, 
$\omega_{QE}=\sqrt{q^2+m_N^2}-m_N$, 
and below it. 
As the CM angle runs from 0 to 180 degrees, for
this kinematics the Lab angle starts growing, reaches a maximum and
then decreases. Therefore, for a given emission angle in the Lab system,
$\theta'_1$, there correspond two angles in the CM, that we denote
$(\theta''_1)^+$ and $(\theta''_1)^-$. They
differ in the value of the Lab momentum $p'_1$, that is plotted in the lower
panel of Fig.~1. Hence there are two different values of $p'_1$
for a given Lab angle. These two $p'_1$-values obviously
correspond to the two solutions, $(p'_1)^{\pm}$ of energy
conservation, appearing in the sum of the phase-space function in Eqs.~(\ref{integral7Drel},\ref{angular}). The momentum of the second
nucleon, $p'_2$, could be obtained by changing $\cos\theta''_1$ by
$(-\cos\theta''_1)$ in Eq.~(\ref{plab}). Therefore the range of
values it takes is the same as $p'_1$.

\subsection{Transformation of the angular distribution}

We assume that the angular distribution in the CM frame is independent of
the emission angle, except for Pauli blocking restictions,
\begin{equation}
n''(\theta''_1) = C \Theta(p'_1,p'_2,0,0) \, ,
\end{equation}
where $C$ is a constant that is determined below. The step function ensures
Pauli blocking. The angular
distribution in the Lab system, $n'(\theta'_1)$,
is obtained by imposing conservation of the
number of particles emitted within two corresponding solid angles $d\Omega'_1$ 
and $d\Omega''_1$, in the Lab and the CM systems
\begin{equation}
n'(\theta'_1)d\Omega'_1 = n''(\theta''_1)d\Omega''_1 \,.
\end{equation}
Since the boost conserves the azimuthal angle $d\phi''_1=d\phi'_1$, we
get the well-known transformation expression:
\begin{equation}
n'(\theta'_1)
=
\frac{C  \Theta(p'_1,p'_2,0,0)}{\left|\frac{d\cos\theta'_1}{d\cos\theta''_1}\right|} \,.
\label{transformacion}
\end{equation}
The derivative in the Jacobian is computed by differentiation of
Eq.~(\ref{coslab}) with respect to $\cos\theta''_1$,
and can be written in the form
\begin{equation}
\frac{d\cos\theta'_1}{d\cos\theta''_1}
=
\gamma p''_1
\frac{p'_1-vE'_1\cos\theta'_1}{(p'_1)^2}
\end{equation}
Writing $\gamma$ in the form:
\begin{equation}
\gamma=\frac{E'}{\sqrt{E'{}^2-p'{}^2}} = \frac{E'}{2E''_1}
\end{equation}
we arrive at the following formula
 for the angular distribution in the Lab frame
\begin{equation}
n'(\theta'_1)=
\frac{2E''_1}{E'p''_1}
\frac{(p'_1)^2}{|p'_1-vE'_1\cos\theta'_1|} C \Theta(p'_1,p'_2,0,0) \,.
\label{distribucion}
\end{equation}
Note that this distribution is not unique, because, as shown in
Fig.~1, there may be two different CM angles, and two different values
of $p'_1$ corresponding to the same Lab angle $\theta'_1$. Therefore
the are two possible angular distributions, and the total distribution
is given by the sum of the two,
\begin{equation}
n'(\theta'_1) =
n'_+(\theta'_1) +
n'_-(\theta'_1) \, ,
\end{equation}
where each partial distribution $n'_{\pm}(\theta'_1)$ corresponds to
Eq.~(\ref{distribucion}) using the $(p'_1)^{\pm}$ values, respectively.

\subsection{Equivalence of Lab distributions}

The next step is to compare the functions $n_{\pm}(\theta'_1)\sin\theta'_1$
with the angular distribution $\Phi_{\pm}(\theta'_1)$ computed for nucleons at
rest, $h_1=h_2=0$,  given by Eq.~(\ref{angular})
\begin{equation}
\Phi_{\pm}(\theta'_1)
= \sin\theta'_1
\frac{m_N^2(p'_1)^2\Theta(p'_1,p'_2,0,0)}{|E'_2p'_1-E'_1\np'_2\cdot\hp'_1|} \, ,
\label{phimasmenos}
\end{equation}
where $p'_1=(p'_1)^{\pm}$.
Using
\begin{equation}
\np'_2\cdot\hp'_1 = q\cos\theta'_1-p'_1
\end{equation}
the denominator in Eq.~(\ref{phimasmenos}) can be written as
\begin{eqnarray}
E'_2p'_1-E'_1\np'_2\cdot\hp'_1
&=&
E'p'_1-E'_1q\cos\theta'_1
\nonumber\\
&=&
E'(p'_1-E'_1 v\cos\theta'_1) \,.
\end{eqnarray}
Substituting in Eq.~(\ref{phimasmenos}) we obtain
\begin{equation}
\Phi_{\pm}(\theta'_1)
= \sin\theta'_1
\frac{m_N^2(p'_1)^2\Theta(p'_1,p'_2,0,0)
}{E'|p'_1-E'_1v\cos\theta'_1|} \,.
\end{equation}
Comparing with Eq.~(\ref{distribucion}), it follows that
\begin{equation}
n'_{\pm}(\theta'_1)\sin\theta'_1 = \Phi_{\pm}(\theta'_1)
\end{equation}
provided that
\begin{equation}
C = \frac{m_N^2}{2} \frac{p''_1}{E''_1}  \,.
\end{equation}

\begin{figure}
\includegraphics[width=8cm, bb=130 240 430 780]{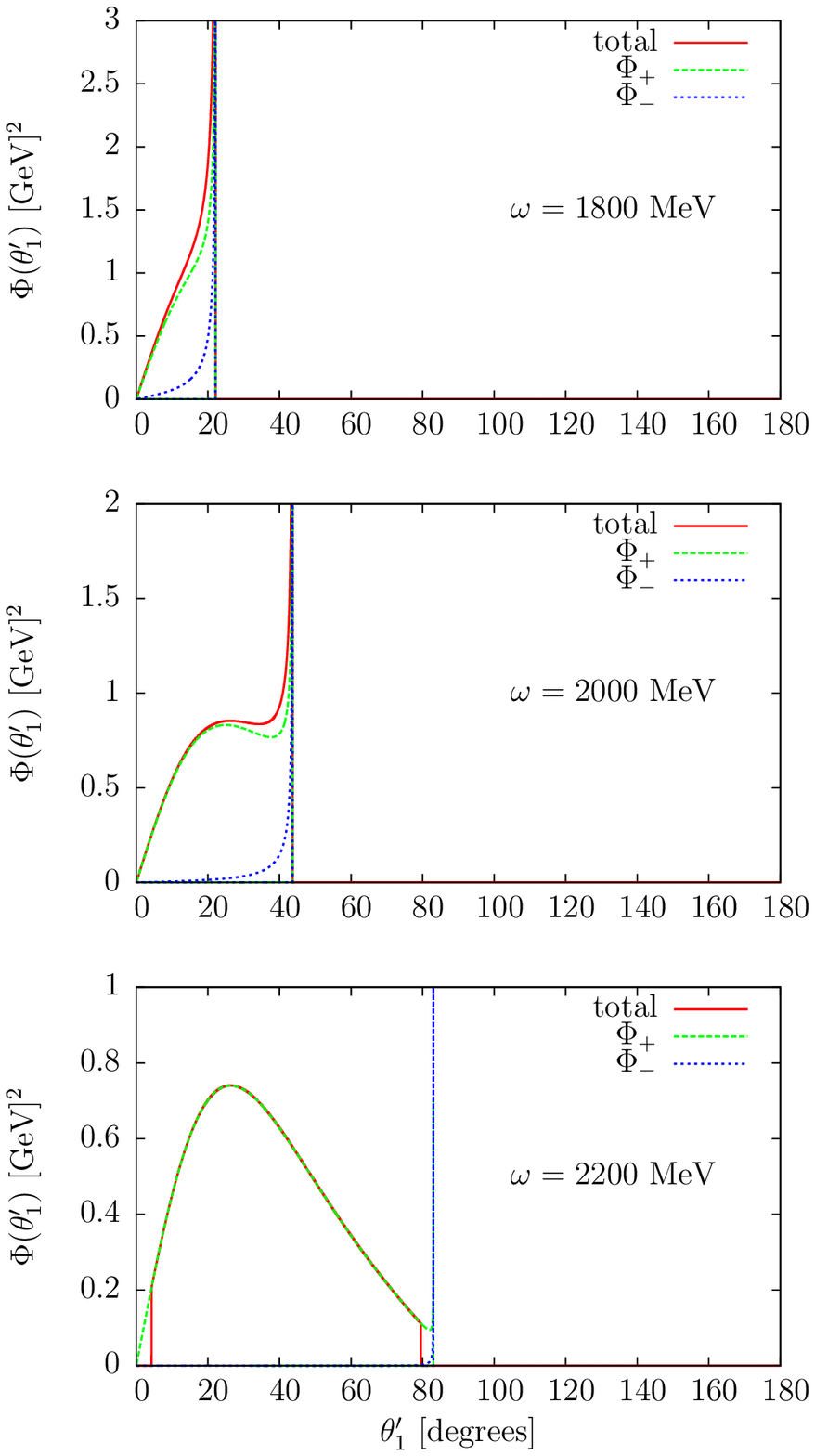}\caption{
\label{fig2}
(Color online)
The two angular distributions $\Phi_{\pm}$ and the total,
in the Lab system, for two-nucleon emission in the frozen nucleon approximation.
The momentum transfer is
$q=3$ GeV/c and three values of $\omega=1800, 2000$
 and 2200 GeV are considered.
 }
\end{figure}

In Fig.~\ref{fig2} we show the two angular distributions
$\Phi_{\pm}(\theta'_1)$ for $q=3$ GeV/c and three values of
$\omega$. We can see that both distributions are zero above a maximum 
 allowed angle in the Lab system.
 Both distributions
present a divergence (they are infinite) at that precise maximum
angle, because the derivative in the denominator of
Eq.~(\ref{transformacion}) is zero at that point. This is
in agreement with our previous work~\cite{Rui14} where we also
demonstrated that the divergence is integrable. The results of
Fig.~\ref{fig2} for the total distribution agree with the findings
of~\cite{Rui14}. In Fig.~\ref{fig2} we have not included Pauli
blocking in the plots of $\Phi_{\pm}$, but it is included in the total
distribution. We see that Pauli blocking only is effective in the last
case, $\omega=2200$ MeV, killing the divergence.

\subsection{Integration in the CM}

The method of the previous section can be reversed by making the
inverse boost from Lab to CM. This allows us to perform the integral over
$\theta'_1$ in Eq.~(\ref{integral7Drel}) using the CM emission angle,
by changing variables $\theta'_1\rightarrow\theta''_1$. Since this is
the inverse transformation applied in the previous sections, the
Jacobian cancels the denominator in Eq.~(\ref{integral7Drel}).

We start by fixing $\nh_1$ and $\nh_2$ and define
the phase-space integral over the final momenta
\begin{eqnarray}
G(\nh_1,\nh_2,q,\omega)
&\equiv&
\int d^3p'_1d^3p'_2
\frac{m_N^2}{E'_1E'_2}
\Theta(p'_1,p'_2,h_1,h_2)
\nonumber\\
&&
\delta^4(H_1+H_2+Q-P'_1-P'_2) \,,
\end{eqnarray}
such that
\begin{equation}
F(q,\omega)=\int d^3h_1 d^3h_2 \frac{m_N^2}{E_1E_2}G(\nh_1,\nh_2,q,\omega) \, .
\end{equation}
We recall from special relativity
that the integral measure
$\int d^3p/E$
is Lorentz invariant because of the result,
\begin{equation}
\int \frac{d^3p}{2E(p)}
=
\int d^4p \, \delta(p^\mu p_\mu -m_N^2)\theta(p^0) \,.
\end{equation}
Then we can write
\begin{eqnarray}
G(\nh_1,\nh_2,q,\omega)
&=&
\int d^3p''_1d^3p''_2
\frac{m_N^2}{E''_1E''_2}
\Theta(p'_1,p'_2,h_1,h_2)
\nonumber\\
&&
\delta^4(H''_1+H''_2+Q''-P''_1-P''_2) \, ,
\end{eqnarray}
where the doubly-primed variables refer to the momenta in the CM frame. The CM is
defined by $\np''=(\nh_1+\nh_2+\nq)''=0$. The step functions, which are not
invariant, must be computed in the Lab system, {\it i.e.,} the momenta
inside the integral have to be transformed back to the Lab system to
compute the argument of the step function. Integrating over $\np''_2$ we obtain
\begin{eqnarray}
G(\nh_1,\nh_2,q,\omega)
&=&
\int d^3p''_1
\delta(E''-E''_1-E''_2)
\nonumber\\
&\times&
\frac{m_N^2}{E''_1E''_2}
\Theta(p'_1,p'_2,h_1,h_2)
\end{eqnarray}
with $\np''_2= -\np''_1$. Therefore, the CM energies satisfy the relationship
$E''_1=E''_2$, and we can write
\begin{eqnarray}
G(\nh_1,\nh_2,q,\omega)
&=&
\int d^3p''_1
\delta(E''-2E''_1)
\nonumber\\
&\times&
\frac{m_N^2}{(E''_1)^2}
\Theta(p'_1,p'_2,h_1,h_2) \,.
\end{eqnarray}
Now we change variables $p''_1\rightarrow E''_1$, and integrate over $E''_1$
using $p''_1dp''_1=E''_1dE''_1$,
\begin{equation}
G(\nh_1,\nh_2,q,\omega)
=
\frac{m_N^2}{2} \frac{p''_1}{E''_1}
\int d\Omega''_1
\Theta(p'_1,p'_2,h_1,h_2) \,.
\end{equation}
The remaining integral of the step function over the emission angles
is in general non-trivial and has to be performed numerically.
If there is no Pauli blocking, the above integral takes its maximum value:
\begin{equation}
G(\nh_1,\nh_2,q,\omega)_{n.p.b}
=
4\pi \frac{m_N^2}{2} \frac{p''_1}{E''_1} \,.
\end{equation}
What remains to be performed is the integral over $\nh_1,\nh_2$, that in
general should be evaluated numerically. However, in
the frozen nucleon approximation one assumes that the integrand
depends very mildly on $\nh_1,\nh_2$, and therefore one can employ this fact to fix the kinematics to the frozen nucleon value, $h_1=h_2=0$.
The phase-space integral in this case is trivial, and takes on the value
\begin{equation}
F(q,\omega)_{n.p.b}
=
4\pi
\left( \frac{4}{3}\pi k_F^3 \right)^2
\frac{m_N^2}{2} \frac{p''_1}{E''_1} \,,
\label{analytical}
\end{equation}
where the ratio $p''_1/E''_1$ in the frozen nucleon approximation is given by
\begin{equation}
 \frac{p''_1}{E''_1}
= \sqrt{1-\frac{4m_N^2}{(2m_N+\omega)^2-q^2}} \,.
\end{equation}
Note that in the asymptotic limit $\omega\rightarrow\infty$,
a constant value is obtained,
\begin{equation}
F(q,\infty)
=
4\pi
\left( \frac{4}{3}\pi k_F^3 \right)^2
\frac{m_N^2}{2} \, .
\end{equation}
This asymptotic limit is in agreement with the one obtained in~\cite{Rui14}
by integration in the Lab system.

\begin{figure}
\includegraphics[width=8cm, bb=130 590 430 780]{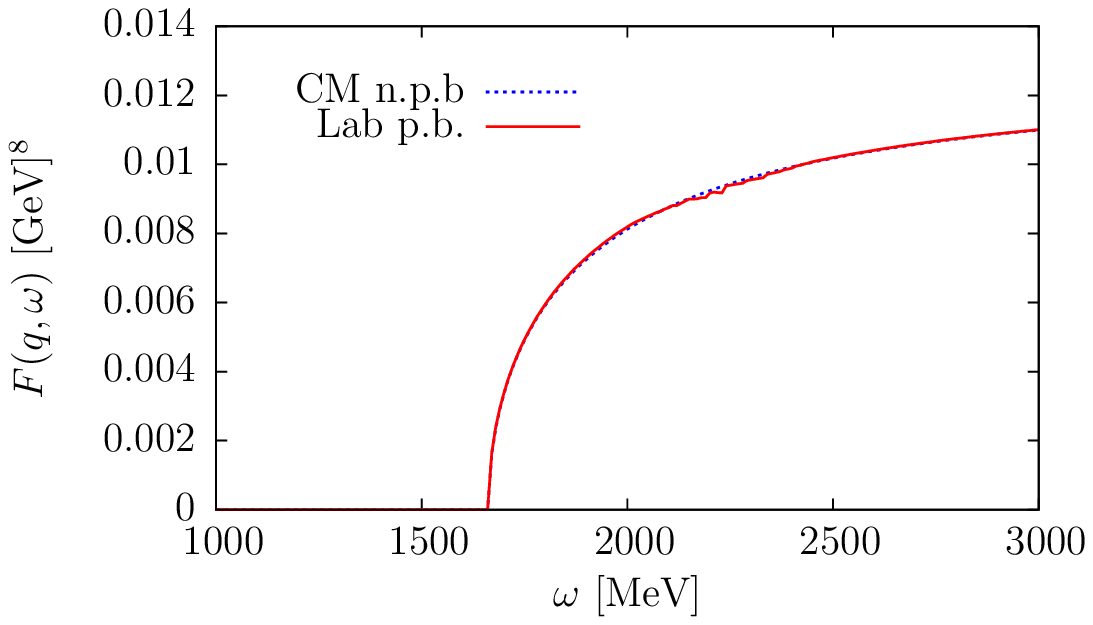}
\caption{
\label{fig3}
(Color online) Phase-space function in the frozen nucleon approximation for
$q=3$ GeV/c, computed in the CM using the analytic formula without
Pauli blocking (n.p.b.), and computed numerically in the Lab system
including Pauli blocking (p.b.).  }
\end{figure}

As an example, we show in Fig.~\ref{fig3} the phase-space function
$F(q,\omega)$ for $q=3$ GeV/c,
computed using the analytic formula without Pauli
blocking, Eq.~(\ref{analytical}), and by numerical integration in the
Lab frame using the method of \cite{Rui14} with Pauli blocking. Both
results agree except in the small region around the quasielastic peak,
where Pauli blocking produces the very small difference seen between the two results; there the Pauli-blocked function $F(q,\omega)$  is slightly below the analytic result.

\section{Conclusions and perspectives}

In this work we have analyzed the angular distribution of 2p-2h
final states in the relativistic Fermi gas, finding the connections
between the CM and Lab systems. Theoretical calculations of
many-particle emission in neutrino and electron scattering usually
rely on the Lab frame to be the most appropriate to perform the
calculations, since the Fermi gas state description is simpler,
mainly because Pauli blocking necessarily has to be checked in the
Lab system where the initial nucleons are below the Fermi surface.
However the description of the 2p angular distribution is simpler in
the CM frame, where the angular dependence is isotropic, if no Pauli
blocking is assumed.

On the contrary, the phase-space integral in the Lab system has the
difficulty that the angular distribution has a singularity at the
maximum allowed angle. The integration of this singularity in the Lab
system was made in our previous work~\cite{Rui14}. Here we have
studied the alternative method of performing the angular integral in
the CM frame, where the angular dependence is trivial. We show that
such an integral can be solved analytically in the absence of Pauli
blocking.

Of interest for the neutrino scattering data analysis, we have shown
that the algorithms used in Monte Carlo event generators produce 2p
angular distributions that are in agreement with the theoretical
calculations in the Lab system if the nuclear current is disregarded.

We have considered the angular distribution coming from phase
space alone. In a complete calculation one is involved with the
interaction between the two nucleons and the lepton that introduces an
additional angular dependence which needs to be evaluated to correctly
describe the events. A proper model of 2p-2h emission requires at
least the introduction of meson-exchange currents, or nuclear
correlations~\cite{DePace03,Ama10a}. Work along these lines is in
progress.

Finally, the integration method proposed here could also be used to
compute the 2p-2h hadronic tensor in Eq.~(\ref{hadronic2}) as an
alternative procedure to the common Lab frame calculations. Comparisons of
the two methods would be of interest because neither of them presents clear numerical
advantages. Although angular integration in
the CM frame allows one to avoid the divergence arising in the Lab frame, it
introduces the difficulty of having to perform a different boost
inside the integral for each pair of holes $(\nh_1,\nh_2)$.

\section*{Acknowledgments}
This work was supported by DGI (Spain): FIS2011-24149 and
FIS2011-28738-C02-01, by the Junta de Andaluc\'{\i}a (FQM-225 and
FQM-160), by the Spanish Consolider-Ingenio 2010 programmed CPAN,
by U.S.  Department of Energy under cooperative agreement
DE-FC02-94ER40818 (TWD) and by INFN under project MANYBODY (MBB).
C.A. is supported by a CPAN postdoctoral contract.


\end{document}